\documentclass[]{spie}  %>>> use for US letter paper
\usepackage{amsmath,amsfonts,amssymb}
\usepackage{graphicx}
\usepackage[colorlinks=true, allcolors=blue]{hyperref}

\newcommand{\Namakanui}{N\=amakanaui}

\title{Implementing Remote Observing at the JCMT}

\author[a]{Harriet Parsons}
\author[a]{Jessica Dempsey}
\author[a]{Dan Bintley}
\author[a]{Craig Walther}
\author[a]{Sarah Graves}
\author[a]{William Stahm}
\author[a]{Maren Purves}
\author[a]{Kevin Silva}
\author[a]{Alexis Acohido}
\author[a]{Graham Bell}
\author[a]{Ryan Berthold}
\author[a]{Jamie Cookson}
\author[a]{Vernon DeMattos}
\author[a]{Devin Estrada}
\author[a]{Miriam Fuchs}
\author[a]{David Fuselier}
\author[a]{Paul Ho}
\author[a]{John Kuroda}
\author[a]{Shaoliang Li}
\author[a]{Steven Mairs}
\author[a]{Mark Rawlings}

\affil[a]{East Asian Observatory, James Clerk Maxwell Telescope, 660 N. A'oh\={o}k\={u} Place, Hilo, Hawai`i, 96720, USA}

\authorinfo{Further author information: Send correspondence to H. Parsons\\H. Parsons: E-mail: h.parsons@eaobservatory.org, Telephone: 1 808 969 6520}

\begin{document} 
\maketitle

\begin{abstract}
The James Clerk Maxwell Telescope (JCMT) is the largest single dish telescope in the world focused on sub-millimeter astronomy - and it remains at the forefront of sub-millimeter discovery space. JCMT continues its push for higher efficiency and greater science impact with a switch to fully remote operation. This switch to remote operations occurred on November 1st 2019.  The switch to remote operations should be recognized to be part of a decade long process involving incremental changes leading to Extended Observing - observing beyond the classical night shift - and eventually to full remote operations. The success of Remote Observing is indicated in the number of productive hours and continued low fault rate from before and after the switch.

\end{abstract}


\section{INTRODUCTION}
\label{sec:intro}

The James Clerk Maxwell Telescope, JCMT, is a 15-m single dish sub-millimeter telescope located on Maunakea in Hawai`i. Since operations began in 1987 the JCMT has made major contributions to astronomy\cite{JCMT-Science-2017RSOS....470754R} in a number of fields, sub-millimeter galaxies \cite{SMGs-1997ApJ...490L...5S,SMGs-1998Natur.394..241H,SMGs-1998Natur.394..248B,SMGs-2013MNRAS.432...53G} , star formation\cite{GBS-2013MNRAS.429L..10H,BISTRO-2017ApJ...842...66W,Transient-2017ApJ...849...43H}, Debris Disks\cite{DebrisDisks-1998Natur.392..788H}, evolved stars\cite{Betelgeuse-2020ApJ...897L...9D}, and most recently detected the molecule phosphine on our neighbouring planet Venus\cite{Science-Venus-2020NatAs.tmp..178G}. As well as remaining a high impact single dish sub-millimeter facility\cite{Crabtree-publications-2018SPIE10704E..0SC}, JCMT with its VLBI capabilities also contributed to the first image of a Black Hole as part of the Event Horizon Telescope \cite{Science-EHT-2019ApJ...875L...1E,EHT-instruments-2019ApJ...875L...2E}. 

Currently the JCMT operates with three main instruments: the SCUBA-2 continuum camera\cite{Holland2013}, which can be used with POL-2 (a front-end linear polarizer\cite{Friberg2018}); HARP\cite{Buckle2009}, a 16 receptor heterodyne instrument operating at 345\,GHz; and \Namakanui, a new heterodyne instrument at the JCMT undergoing commissioning with capabilities at 86GHz, 230GHz and 345GHz. The observatory also has made efforts to push forward on future instrumentation plans\cite{future-instrument-jcmt-2016SPIE.9908E..07D}. Most notably, a third generation dedicated 850 micron camera, guaranteeing an increase of at least an order of magnitude in mapping speed relative to SCUBA-2, with intrinsic polarization capabilities. This is needed to ensure that in the era of multi messenger astronomy JCMT continues to contributes to cutting edge research\cite{EAO-futures-2020arXiv200203100S,EAO-Futures-2020arXiv200105753F,EAO-Futures-2020arXiv200107447W,EAO-Futures-2019arXiv191201620M}.

In recent times, many facilities have shifted to remote operations. For examples one needs to look no further than the Maunakea Observatories to see this change in operational policy and attitude. In 2008, the UH88 was the first telescope on Maunakea to switch to full remote observing followed by UKIRT in 2010 \cite{UKIRTremote-2016SPIE.9913E..32H}, CFHT in 2011 \cite{CFHT-remote-2011tfa..confE...8G,CFHTremote-2014ASPC..485...53V} and Gemini North in 2015/2016 (known as Base Facility Operation, BFO, \cite{GeminiRemote-2016SPIE.9911E..1QN}). Keck has also begun investigating the the process to switch over to Unattended Night Operations (UNO\cite{KeckUNO-2018SPIE10705E..06G,ProjectManage-Remote-2018SPIE10705E..1BG}).

The impetus to shift to remote operations for many facilities, including JCMT, is primarily driven by cost. This relieves the funding pressures from agencies supporting visiting researchers who were required to provide - primarily for safety - a second person at the summit. Additionally, the summit of Maunakea is located at 14,000'. At this elevation, the atmospheric pressure is about 40\% less than at sea level. These conditions are mentally and physically taxing. Locating telescope operating staff at the sea-level facility is safer, and allows for clearer decision-making capacity as well as enabling in person communication between the TSSs (Telescope System Specialist, also referred to as Telescope Operator) and support staff. 

Our aim in this paper is to outline the path to Remote Observing that was taken at the JCMT. One year later the JCMT continues to see a consistent low fault rate and high number of productive hours on sky each night. In addition, the agility that this switch to remote operations has provided has enabled successful operations during the current ongoing global pandemic with scientific productivity remaining strong.

\section{IMPLEMENTATION: A DECADE IN THE MAKING}
\label{sec:implementation}

\subsection{A Decade in the making}
The switch to remote operations of the JCMT in November 2019 (operations of the un-staffed summit facility from Hilo) at the JCMT should be recognized to be part of a larger framework of a deliberate long term shift in attitudes and capabilities at the facility over the past decade. Under the operations of the Joint Astronomy Centre\footnote{Historically the Joint Astronomy Centre (JAC) operated the United Kingdom Infra-Red Telescope (UKIRT) until October 31st 2014 and the James Clerk Maxwell Telescope (JCMT) until March 1st 2015. Legal ownership of the JCMT then transferred from STFC to the University of Hawaii and the telescope is being operated by East Asian Observatory in partnership with the UK and Canadian research communities.  } UKIRT successfully switched to remote operations in December 2010 \cite{UKIRTremote-2016SPIE.9913E..32H}. This was a ``soft switch" to remote operations thanks in part to the JCMT night time operators that could provide support and assistance as needed to UKIRT - verifying ice/weather conditions and providing moderate on-site support due to the proximity of the facilities and shared infrastructure. This switch to remote operations at UKIRT undertaken by JAC staff meant that the JCMT benefited from existing shared software and policies, and other insights from this process ten years prior.

\subsection{Soft testing under Extended Observing}
The initial steps on the road to Remote Observing was enabling the observatory to execute ``Extended Operations" (EO)\cite{Walther-EO-2014}. JCMT is a sub-millimeter observatory and is able to operate during the daytime\footnote{Diurnal degradation in sub-milimeter atmospheric transmission  and  stability,  as  well  as  thermal  deformations  of  the dish means daytime observing is rarely performed and only for specific science cases.}. EO sees the observatory operated beyond the classical 12-13 hour night shift into the morning thus providing the astronomical community additional scientific data prior to the detrimental effects of dish heating and atmospheric sky instability.

Initial work focused on the hardware and software needed to  control  the  roof  and  door closure and telescope drive systems remotely through the JCMT programmable logic controller (PLC)\cite{Walther-EO-2014}. This work was undertaken in August and September of 2013,  with  the  first testing  of  the  procedure  successfully completed in early October that year. Interior and exterior cameras for safety as well as an IR camera to enable the monitoring of clouds were installed. Importantly, an automatic shutdown procedure to close the JCMT was developed that would execute if a software handshake (known as the dead man handshake) was not maintained between the Extended Operator and the telescope control system, no matter what the circumstances\cite{Walther-EO-2014}.

With the work complete, a group of observatory staff volunteered to become the observatory's first Extended Operators.  From late-October through to December 2013, EO  was successfully  conducted  remotely from Hilo between Monday and Thursday when weather conditions were favourable and the daytime technicians and engineering staff were available to provide onsite assistance if any faults occurred. 

After system testing by staff and the hire of two part-time Extended Operators, full seven-day-a-week  EO officially began in January 2014. It is estimated that the increase in observing time from EO is equivalent to two nights per month\footnote{Based on numbers from March 2015 - September 2019 ~22.8 hours of productive time per month (time on science and calibration).}.

On 15 December 2014 the JCMT operators were given the go ahead to open the JCMT doors remotely in an attempt to cool down the telescope prior to observing, increasing telescope efficiency. 

Although work had begun shifting to a model of remote operations, efforts ramped up to the point that the observatory was ready and capable by February 1st 2016 to go fully remote on a partial basis. This effort was made to enable a key staff member to continue to work through pregnancy operating the telescope (which otherwise would not have been advised at high altitude). With the TSS shift model of 5 nights on 10 nights off it meant 33\% time undertaking observations remotely from Hilo. Full summit nighttime operations at the JCMT resumed in September 2016.

\subsection{The transition}

In September 2019, the final effort was made to switch to full Remote Observing with a deadline set of November 1st. Observatory staff lead a comprehensive internal review of all hardware, software systems and safety that would be affected by the change with input invited from \textit{all} staff at the observatory. At this point, almost all of the necessary technical changes were in place and had been in use for a number of years.

The focus of this review was risk assessment and mitigation. Risks reviewed included increased risk of damage to the unattended facility, the potential increase in instrument down-time and the possible reduction in the science data quality. The primary risk from Remote Operations came from the drop in summit facility occupancy\footnote{It is estimated that staff are at JCMT for 20\% of the time (in any week) compared to 70\% before Remote Observing. This is based on classical 13 hour night shifts 7 days a week and day time engineering work Monday-Thursday. This estimate does not account for daytime or VLBI observing.} with support staff a 2 hour journey away from the facility. To safely implement fully remote operations required establishing and then implementing new procedures for both the engineering daycrew and for the TSSs. 
The major components of this work is discussed below.

\subsubsection{System upgrades}

Much of the critical facility work to support remote operation was done prior to 2014, for remote Extended Observing. Essential equipment and controllers were already on both UPS and remote power switches or network power switches. 

Final steps in the transition to remote operation included placing all Linux and VxWorks boxes on remote power switches, with their consoles also accessible remotely. Some older PCs that had required manual restarts were upgraded or replaced. Additionally the Secondary Mirror Unit (SMU) electronics, and Tertiary Mirror Unit (TMU) controller were placed on network power switches.

The JCMT's instrumentation suite of critical components could be individually power cycled via network switches: motor controllers, bias controls, CAN nodes, cryostat heaters, amplifiers etc. Upgrading HARP power supplies was a non trivial task, requiring rewiring so that the various nodes could be power cycled remotely. Network power switches are critical tools for remote operation, as they allow equipment revival and troubleshooting remotely, though the failure of the switches themselves is an added risk, and switch performance must be carefully monitored.

\subsubsection{Instrumentation}

Long before the plan for full-time Remote Observing, all JCMT instruments had been capable of remote operation. The tasks from setting up an instrument or tuning a receiver, to loading an observation recipe while running the JCMT observing and telescope control software could equally be done in the JCMT control room or from a remote computer logged-in to the JCMT network. The establishment of JROC for remote Extended Observing ensured that the necessary computer workstations and network infrastructure was in place in Hilo and that all of the software and complex interconnections worked without hitch.

JCMT instruments are also designed to be safe in the event of unexpected warm-ups or power outages. The risk mitigation for instrumentation was therefore to ensure the minimum loss of time from the inevitable minor faults, that in the past, the TSS could fix. Typically, this involves faults affecting the Closed Cycle Coolers (CCC) used maintain the 4K internal temperatures necessary for the instrument operation. 

All JCMT instruments require the continuous running of compressors (for their CCC cryogenics) and recirculating water chillers to cool the compressors. JCMT does not currently have the capacity to put compressors or water chillers on UPS. Therefore the systems rely upon the water chillers restarting automatically after any power outage or power glitch (which are very frequent at the summit). The startup of each chiller is staggered (using delay relays), to avoid overloading JCMT's power circuits and each chiller will then, after another short delay, start its connected compressor. 

This system works well but is not flawless. Occasionally a chiller does not restart due to a system fault or in the case of SCUBA-2 which uses Pulse Tube Coolers (PTCs), the compressor motor can restart but with no 'heartbeat' and therefore no cooling\cite{Cookson2016-sc2-cryogens}.

Typically the TSS would simply power cycle the affected chiller/compressor and that would fix the cooling fault. Also common are failures of the water chillers themselves or water/glycol leaks. A spare chiller (and also spare compressors) is kept on standby for each instrument and JCMT TSSs were well versed in swapping to the spare when necessary, often with no loss of use of the instrument.

At JCMT, the PLC provides an audible alarm and a display that indicates if any instrument chiller and compressor is off or in a fault state. Instrument temperatures are also monitored, and will generate email alarms when outside limits. Similar warnings appear on the JCMT Tonight web page and in the Instrument setup GUI for the relevant Instrument.

\subsubsection{SCUBA-2 cryogenic system}

SCUBA-2 is a highly sensitive sub-millimeter continuum camera operating at 450 and 850 microns with over 10,000 bolometers cooled below  70\,mK\cite{Holland2013}. The SCUBA-2 cooling system\cite{Cookson2016-sc2-cryogens} is a dilution refrigerator with three PTCs providing the various cooling stages within the cryostat. The room temperature Gas Handling System (GHS) that circulates $^{3}$He through the dilution fridge is also designed to recover the mixture safely should the instrument warm up unexpectedly. 

Over the years of SCUBA-2 use, many improvements to the GHS coupled with careful maintenance of the pumps and seals have kept the instrument functioning safely and recovering the mixture in the event of serious faults or total loss of power at JCMT. However, in many cases the TSS would assist in fixing minor faults before the automated mixture recovery could start. In the case of a more serious fault that could not be fixed by the Operator, they would assist in recovering the mixture manually, to separate the $^{3}$He from $^{4}$He. This recovery task is key to reducing the time to cool the instrument again, once the fault is fixed.

On switching to full remote operations the goal has been to provide the TSS with the equivalent monitoring of SCUBA-2's closed Cycle Cooling health as at the summit, together with additional tools that enable remote power cycling for the SCUBA-2 recirculating chillers, which in turn provides remote control of the PTC compressors. 

The TSS is made aware when the 4K and 40K stages are warming (which usually indicates that one or more PTC has stopped) and has the option to power cycle an individual or all three chiller and PTC compressor combinations.

The ability to remotely control the power of the SCUBA-2 chillers/PTC compressors has been used effectively in the past 12 months to restore SCUBA-2 to full operational health. However, this is not enough. To make the \textit{correct} decision, the TSS ideally needs additional information about any fault state of the chillers or the compressors before making the decision to power cycle. Work over the past few months has focused on being able to reliably monitor the pulse tube heartbeat without impacting PTC performance. The ultimate aim will be to provide this health indicator, together with the chiller fault state, remotely to the Operator.

Additionally, from the start of full-time remote operations the TSS has been provided with a GUI representation of the SCUBA-2 dilution fridge GHS control front panel. This provides a way to remotely stop and start the dilution fridge and recover the mixture if need.

Previously TSSs were trained to use the physical GHS control front panel to recover the mixture, or if in doubt to hit the ``RESET" button to stop all pumps and close valves. RESET leaves the instrument in a safe state from which it can warm-up unattended, with no risk of high or over-pressure events that can result in loss of $^{3}$He.

A safer version of the remote GHS panel/GUI that provides the Operator initially with just a remote RESET button for the dilution fridge and hides the other valves and controls unless enabled is being worked on.

Overall, remote operation has cost no additional loss of SCUBA-2 available observing time in the past 12 months and the additional safety features added to the GHS have reduced the risk of possible loss of the dilution fridge mixture.

\subsubsection{HARP cryogenic system}

HARP is a unique world-leading 16 receptor Heterodyne Array Instrument. HARP is cooled by two Closed Cycle Coolers. A CTI GM coldhead cools the cold optics while a Sumitomo Daikin coldhead that includes a JT $^{4}$He expansion stage cools the mixer array block and LNAs.

The Daikin compressors have extensive pressure interlocks that make restarting after a power outage more difficult. The JT has a micrometer needle valve assembly and depending upon system temperatures, it is sometimes necessary to open the JT valve when cooling back down, else the Daikin compressor will trip and shut down.

So far no changes have been undertaken for remote operations. Remote power cycling of the chiller/compressor pairs together with additional monitoring of chiller fault states is planned. In addition, motorising the JT needle valve is under consideration.

\subsubsection{N\=amakanaui cryogenic system}

\Namakanui, JCMT's three receiver cartridge Heterodyne Instrument, is cooled by a three stage Sumitomo GM coldhead. This has proved quite robust in the past 12 months and restarts automatically after power outages. Like with HARP, \Namakanui\ will have remote power cycling of the chiller/compressor installed.

\Namakanui\ is also a good candidate to add a second water chiller that would switch-in automatically in the event of a fault with the running chiller. Similar arrangements were implemented at UKIRT. 

\subsubsection{The roof and door system}

In 2013, after substantial work, it became possible to remotely control the JCMT's roof and door system, thus paving the way for EO\cite{Walther-EO-2014}.  As part of the shift to remote operations it was recognized that an extensive overhaul to the roof and door electrical and pneumatic systems was required. The main driver behind this work is the risk from the Roof and Doors failing to close in bad weather. The engineering work required was broken up into a number of stages that could be completed in a standard workday - lessening the impact to night time operations. This work was undertaken in October through to December 2019, and completed in October 2020. 

\subsubsection{Weather}

In situ weather observations are understandably a key component for the preferred method of making weather assessments for an observatory facility. At the JCMT, visual and IR cameras were installed to assist with the monitoring of conditions throughout an observing shift. In particular, IR cameras for early fog and high humidity have been invaluable, providing early detection of degrading weather conditions. With the IR cameras a TSS is able to see fog when it approaches the summit ridge giving adequate time to close the JCMT roof and doors before inclement weather arrives. Such is the importance of the visual and IR cameras that at JROC the remote cameras are on a separate display for easy monitoring. The cameras also provide assistance with facility safety, enabling operators to check areas internal to the observatory once the daytime engineering team have departed. In addition to cameras, all Maunakea TSSs (both remote and on-site staff) share real-time information via a Slack channel which can also help assist remote operators\cite{2018SPIE10704E..1RD}, although as more and more of the telescope facilities on Maunakea go to remote operations - on-site condition reporting may become obsolete.

To mitigate the risk of sudden weather degradation that might not be immediately apparent to the remote TSS, the upper limits on the humidity and wind speeds that would trigger closure of the facility were made more conservative. An operator may only have the facility open when conditions fall within these limits. 

The only necessary function that is not possible to do remotely is a facility inspection following a snowfall. This has required the TSS to travel to the summit twice since the start of remote operations. Daytime engineering staff along with Maunakea ranger information and on-site operators at other Maunakea facilities have provided useful information.

More conservative upper limits on safe observing conditions has resulted in facility closure under Remote Observing guidelines that likely would have not occurred with a TSS at the summit. The reduced humidity limit has resulted in roughly six hours of additional closure time during the year of remote operation. The lower wind speed limit has reduced operational hours on sky on eight nights during the past twelve months. The hours lost are not significant over such a time period and reasonable given that facility safety is the priority. The limits will undergo a comprehensive review and perhaps be increased following planned upgrades to the JCMT weather station.

\subsubsection{Science quality}

Prior to Remote Observing the JCMT was staffed each night by a TSS and a visiting scientist. During summit operations a visiting scientist was a critical safety element to night time operations - required due to the remote nature and high altitude risks associated with observing at 14,000'. Visiting scientists were additionally asked to contribute to high level data checks and science program decisions. Over time, the software tools and resources available to both summit observers and remote PIs/COIs have become more sophisticated, making the visiting observer's role of real-time data checking mostly redundant.

\begin{figure}
    \centering
    \includegraphics[width=0.9\linewidth]{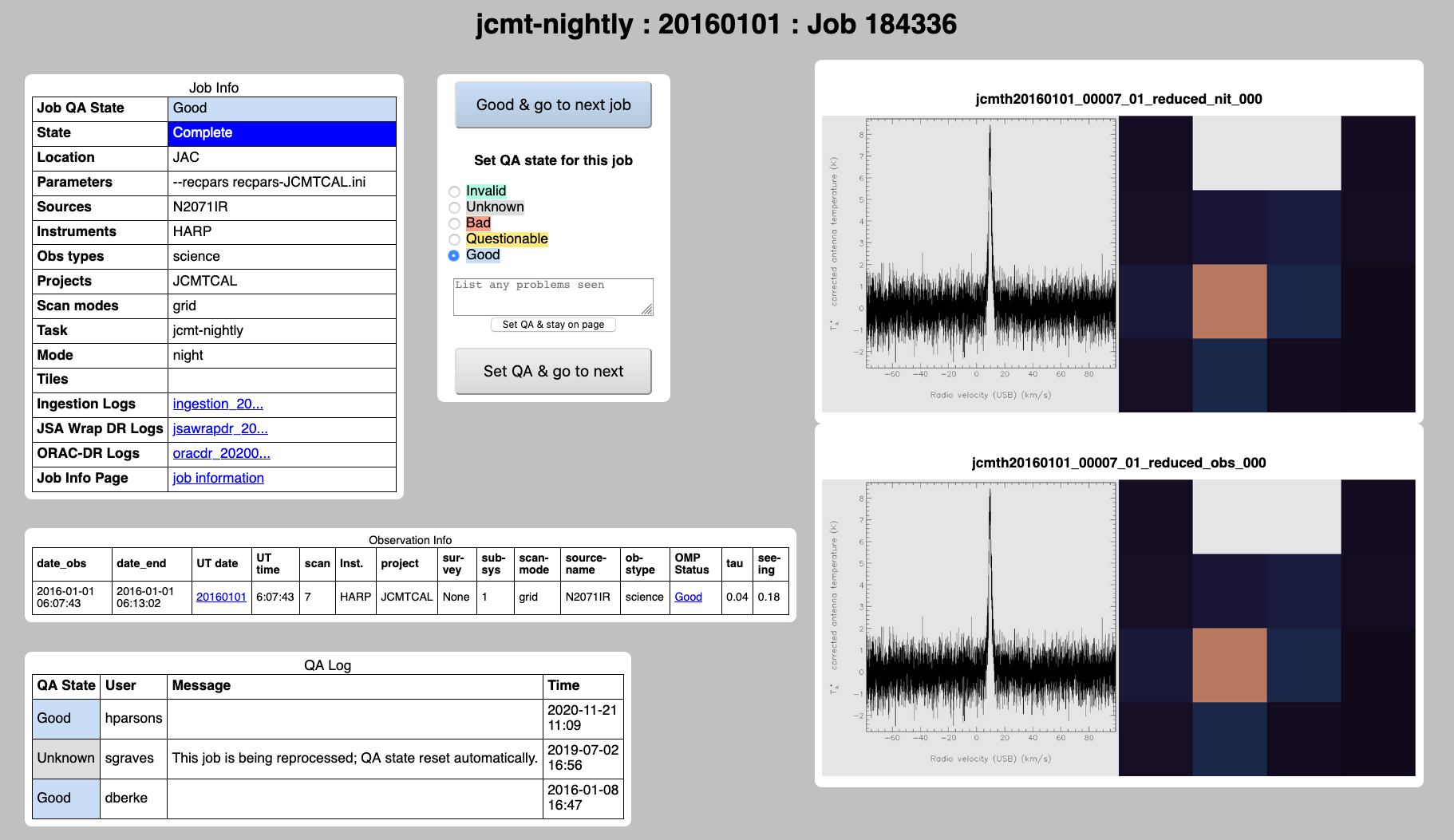}
    \caption{Example of a science observation reduction page in the JSA processing tracking system. A JCMT calibration observation of N207IR taken by the instrument HARP in 2016 is shown.}
    \label{fig:jsa_proc}
\end{figure}

The tools referred to include i) the ORAC-DR\cite{Jenness2015-ORACDR} data driven pipeline that is used by both summit operators and PI. At the summit, operators run the pipeline automatically producing a quick snapshot of reduced data taken on a night. The observatory subsequently runs a more comprehensive data reduction using ORAC-DR to produce products that are sent to the ii) JCMT Science Archive (JSA)\cite{Economou2015A-JSA} at CADC (Canadian Astronomy Data Centre\footnote{\url{https://www.cadc-ccda.hia-iha.nrc-cnrc.gc.ca/en/jcmt/}}) that hosts all JCMT data products and may be accessed by the PI the next day. iii) the OMP system that facilitates the recording of initial data flagging in the system (data flagged as ``questionable'' are flagged for follow up by a staff scientist supporting the related program). iv) the JSA processing tracking system\footnote{\url{https://github.com/eaobservatory/jsa_proc}} - created in 2014 - that provides a way to track the progress of data reductions performed at the observatory prior to ingestion at CADC. This database is of value to the archivists, and also for science support staff to review data quality (see Figure \ref{fig:jsa_proc}).

With the above software tools in place a PI/COI may still connect to the observatory during a night of operations. A new features enables Science users to be automatically emailed when their observations are underway on a given night, and are invited to join the TSS in real-time over online Zoom sessions should they choose.

\subsubsection{Community engagement}

The value of visiting scientists - members of the telescope community - at an observatory should not be underestimated. More prolific and robust science is produced when the authors have a solid understanding of the instrumentation, observing modes, data collection and reduction methods and these are most easily communicated in situ and in person.

Anecdotal evidence suggests that when users have at least one visit to the telescope they become more engaged users. During a visit they can interact with staff improving their understanding on data quality, instrument capabilities, and ultimately write better future proposals, and have a shorter turnaround between data collection and publication. 

Although stalled due to COVID-19, the observatory will launch a community engagement program to support early career astronomers in the future. Under this program, early career astronomers will be invited to visiting the Hilo, sea-level- observatory facility for a period of a few weeks, given the opportunity to visit the telescope and to collaborate with science staff. In the interim JCMT staff will be providing virtual workshops on a range of subjects for new and interested JCMT users starting in early December 2020.

\subsection{An operational model designed for Remote Observing}
\label{section:operations-model}

Key features of the existing JCMT operational model gave confidence in the observatory's ability to continue to produce efficient high quality science after switching to Remote Observing.

The JCMT operates a Flexible Queue Observing Policy (sometimes referred to as ``dynamic scheduling")\cite{FlexibleObserving-2002SPIE.4844...86R,Queue-Scheduling-Tilanus2000,Dempsey-flexible-classical2014} with programs selected for observing based on current source availability, scientific ranking, weather and instrument availability. This policy combined with the OMP (Observation Management Project) software ``a comprehensive web-based feedback system that provides a communication gateway between PIs, observers and staff."\cite{Economou2002-Felixble-observing-software,Delorey2004-OMP}), ensured observations could be out without observers ever coming to the telescope. This reduced the need for observers at the telescope and eased the transition to Remote Observing. 

Efficient use of observatory time pre- and post switch to Remote Operations has been provided by the remotely operated Tertiary Mirror Unit (TMU), enabling the rapid switch between instruments. Over the years hardware and software upgrades have ensured that this key component to observing performs reliably. Switching instruments takes a few seconds - albeit calibration overheads (pointing etc) means that there is minimal cost to switching instruments - mitigating fault loss time and enabling efficient use of telescope time. 

Manual night time cryogenic refills are obsolete at the JCMT resulting from upgrades to instrumentation over the years. Currently all instruments in operation at the JCMT utilize closed-cycle cooling systems and none require cryogenically cooled calibration loads for science operation, easing the switch to Remote Observing. 

Finally it is noted that the JCMT TSSs are trained to a high level of expertise, such that faults that do arise rarely require additional staff input in the moment, and the assessment of scientific data quality happens on-the-fly with timely fault reporting and resolution via the OMP system.

\section{REMOTE OPERATIONS - THE PROOF IS IN THE NUMBERS}

\begin{figure}
    \centering
    \includegraphics[width=0.6\linewidth]{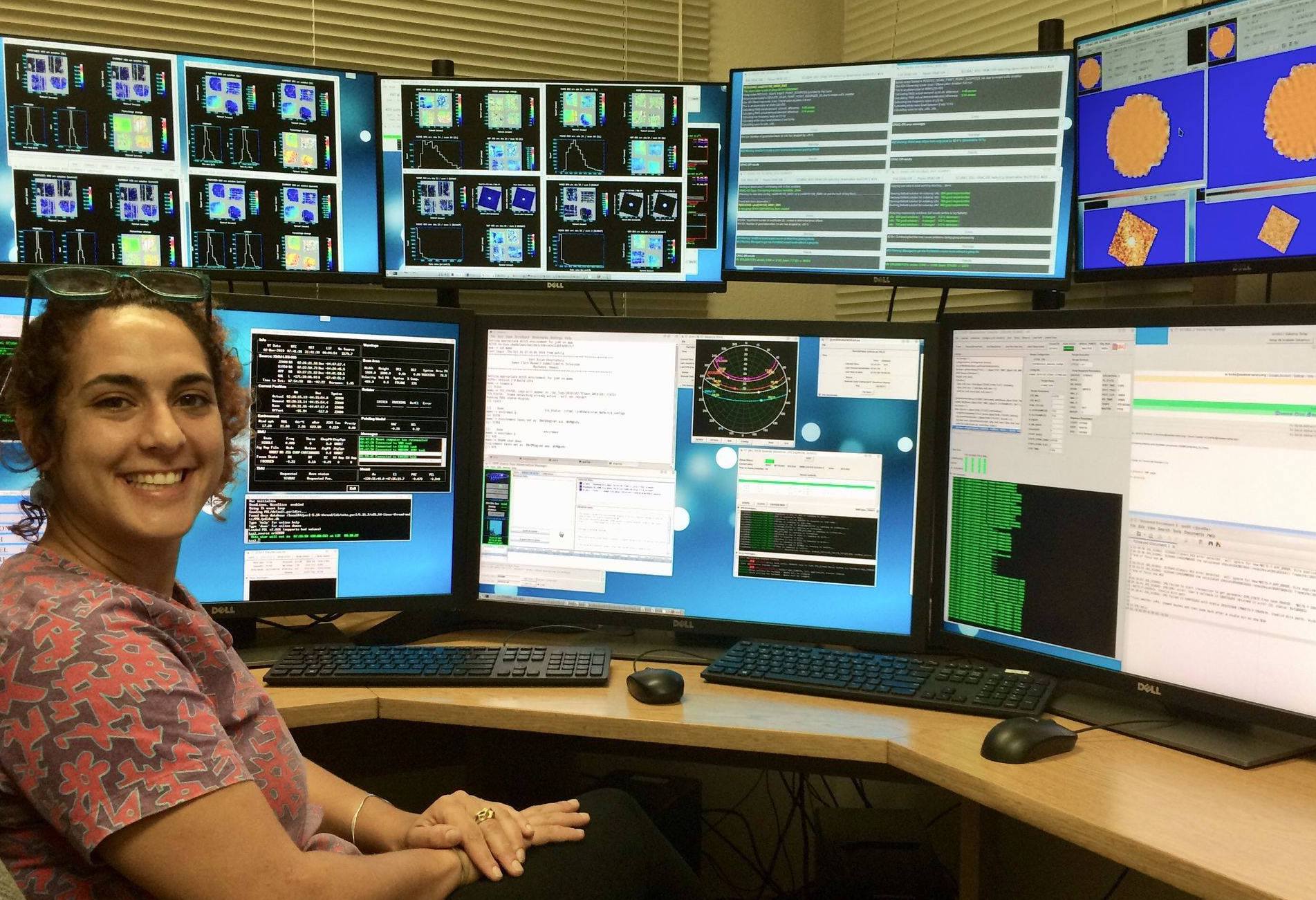}
    \caption{First night of Remote Operations on November 1st 2019 from the JCMT Remote Control room (JROC) by JCMT TSS Mimi Fuchs.}
    \label{fig:FirstNightJROC}
\end{figure}

On November 1st 2019 Remote Observing began full time at the JCMT (see Figure \ref{fig:FirstNightJROC}). Remote operations have continued  for over 12 months now allowing for an assessment of the success of the switch. A comparison of the pre- and post-remote operational metrics including the fault rate and productive time on-sky provides a measure of the success of the JCMT remote operation. \footnote{Since March 2019 the JCMT time accounting software tracks the number of hours spent on sky on a per-shift basis - Nighttime, Daytime and EO - previously EO hours were not tracked to the same level of detail.}. For this comparison the observatory performance pre-remote operations (between March and June 2019\footnote{The Summer 2019 operational hiatus resulting from restricted access to Maunakea began July 2019 and an extremely wet observing season Autumn/Winter 2019 and second operational hiatus (March - May 2020) due to COVID-19.}),  is compared to the operational performance, post-remote operations (between June and October 2020). 

\begin{figure}
    \centering
    \includegraphics[width=0.9\linewidth]{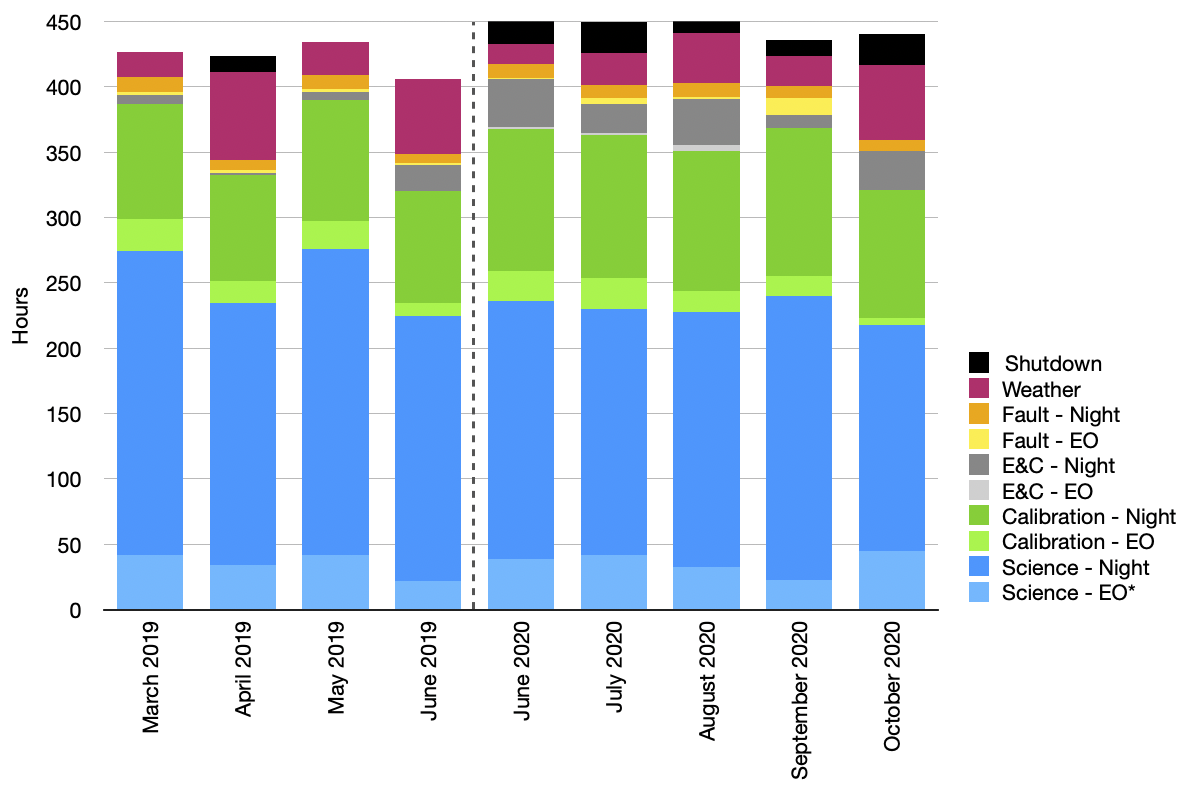}
    \caption{Observatory hours broken down by category. Science, Calibration, Engineering and Commissioning(E\&C), Fault, Weather (time when observatory was closed due to weather), and Shutdown (time when observatory was closed). The months displayed are 1) four months (March - June 2019) prior to shutdown due to the restricted access on Maunakea when the JCMT was operated from the summit. 2) the five months after the return to operations (June - October 2020) from global pandemic shutdown when the JCMT was operated remotely. The time gap and switch to Remote Operations are indicated by the dashed vertical line.}
    \label{fig:hours-EO-night}
\end{figure}

A break down in the hours spent during night time an EO by month is provided in Figure \ref{fig:hours-EO-night}. A summary table, this time looking at per night averages, is also provided (Table \ref{tab:2019-2020-performance-comparison}). Looking at the total number of productive time (as provided in Table \ref{tab:2019-2020-performance-comparison}), on average the observatory is now obtaining \textit{more} productive hours on average since the switch the remote operations (despite the slight increase in fault rate). The productive hours on the telescope have increased by approximately one hour per night which is inline with the time taken for a TSS and observer to take a round-trip between Hale Pohaku (the base accommodation facility for observatory staff working on Maunakea) and the JCMT. The other notable change in Table \ref{tab:2019-2020-performance-comparison} is an increase in calibration and E\&C time. This is the result of commissioning the new heterodyne instrument \Namakanui, which saw first light on the telescope in October 2019.

\begin{table}[ht]
\caption{Summary table of JCMT performance before and after the switch to Remote Observing, for a four month window in 2019 and a five month window in 2020. Hours per night account for closures due to observatory shutdown and weather. Productive time includes hours spent on Science, Calibration and Commissioning. \dag March-June 2019: after an update to operational time accounting and prior to restricted Maunakea access. \ddag June-September 2020: after the COVID-19 JCMT operational hiatus. \S E\&C: Engineering and Commissioning. } 
\label{tab:2019-2020-performance-comparison}
\begin{center}       
\begin{tabular}{|c|c|c|c|c|} %% this creates two columns
%% |l|l| to left justify each column entry
%% |c|c| to center each column entry
%% use of \rule[]{}{} below opens up each row
\hline
\rule[-1ex]{0pt}{3.5ex}   &  \multicolumn{2}{c|}{\textbf{Average per night (2019)}\dag} &  \multicolumn{2}{c|}{\textbf{Average per night (2020)}\ddag} \\
\hline
\rule[-1ex]{0pt}{3.5ex}   & \textbf{Night Time} & \textbf{EO} & \textbf{Night Time} & \textbf{EO} \\
\hline
\rule[-1ex]{0pt}{3.5ex} Science (hrs) & 8.2 & 1.3 & 7.42 & 1.1   \\
\hline
\rule[-1ex]{0pt}{3.5ex} Calibration (hrs) & 3.3 & 0.7 & 4.1 & 0.6  \\
\hline
\rule[-1ex]{0pt}{3.5ex} E\&C (hrs)\S & 0.3 & 0.0 & 1.0 & 0.1  \\
\hline
\rule[-1ex]{0pt}{3.5ex} \textbf{Productive time (hrs)} & \textbf{11.7} & \textbf{2.2} & \textbf{12.5} &\textbf{ 2.0 } \\
\hline
\rule[-1ex]{0pt}{3.5ex} Fault rate(\%) & 3.0 & 4.8 & 3.1 & 8.7  \\
\hline
\rule[-1ex]{0pt}{3.5ex} Overall fault rate (\%) & \multicolumn{2}{c|}{2.9}  & \multicolumn{2}{c|}{3.4}  \\
\hline
\end{tabular}
\end{center}
\end{table}

The fault rate has remained consistent at or below 3\% following the switch to remote operations and well below the Observatory target of less than 5\%, without a single instance of the TSS requiring access to the facility to resolve a fault in the past 12 months. 

\begin{figure}
    \centering
    \includegraphics[width=0.49\linewidth]{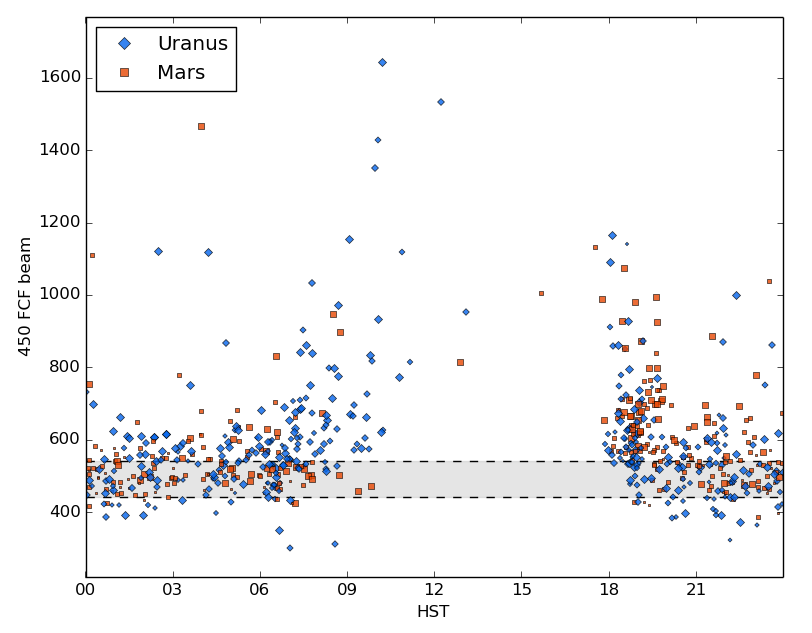}
    \includegraphics[width=0.48\linewidth]{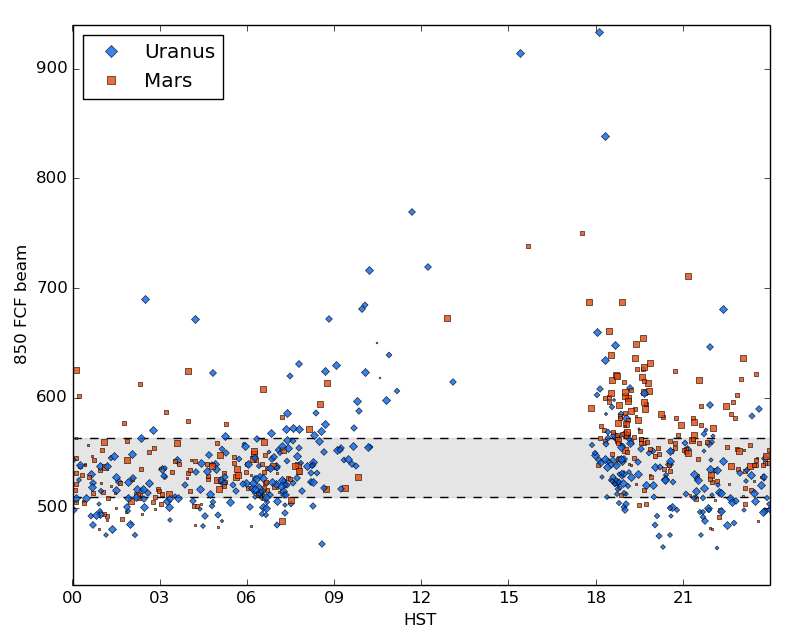}
    \caption{The FCF (Flux Conversion Factor) at 450 and 850 microns as a function of HST (local time) observed in good weather (perceptible water vapour $<$1.75mm). The size of the symbol is proportional to the atmospheric transmission at 225\,GHz.}
    \label{fig:sc2-hst}
\end{figure}

The observatory is now close to approaching the maximum number of productive hours on sky that can be obtained with the current instrument suite. The limiting factor is the diurnal degradation in sub-milimeter atmospheric transmission and stability, as well as thermal deformations of the dish. This can be seen when one considers the  The FCF (Flux Conversion Factor), the value needed to convert data from pW to Jy - as a function of time of day (see Figure \ref{fig:sc2-hst}). The FCF is calculated by observing one of our ‘calibrators’, which are bright point-sources having well-known flux densities. 

\section{AGILITY IN THE FACE OF ADVERSITY}
\label{sec:agility}

At the time of the original submission of this paper the authors' focus was ensuring continued engagement of the JCMT astronomical community with the staff and facility. This is still a long term priority, however in light of the global pandemic it is felt more appropriate to discuss the impact of COVID-19 on Observatory operations.

One of the main drivers behind the switch to remote operations at the JCMT was the ability to operate the telescope remotely without the requirement for a second person at the facility. During summit-based operations, a second person was required to be present with the TSS during observing, due to the inherent risk involved when working at a remote facility at 14,000'. Historically this was fulfilled by a visiting observer with typical observing shifts lasting between 3-6 nights. As of November 1st 2019, a visiting observer was not required and so reducing the international travel from outside of Hawai`i by scientists from East Asia, Canada and the UK (from the JCMT's nominal funding regions). As the global pandemic unfolded this clearly mitigated risk to both local staff, the local community and the visiting observers\footnote{With the switch to Remote Observing on November 1st it is estimated to have reduced visitors at the JCMT by 50 in the first two observing semesters, 18 observing blocks non-UH in 19B, 32 observing blocks non-UH in 20A.}. 

Remote Observing enabled the JCMT to respond fast and efficiently to the pandemic. At the time of the state mandated COVID-19 operational hiatus, the JCMT had been running remote operations for 4.5 months. Once State restrictions on astronomy were lifted, JCMT operations resumed in just two days, and new COVID-19 policies to ensure minimal staff interaction and social distancing could be easily implemented. An additional precaution taken by the observatory was the implementation of a second JCMT Remote Observing Control room  (JROC2). The expansion to provide a JROC2 further minimized staff contact with the TSS observing from JROC and the EO operator from JROC2.

\section{SUMMARY}
\label{conclusion}

When the JCMT began full remote operations of the JCMT on November 1st 2019, it was after a decade long journey by staff to ensure the observatory's readiness. The goal for the JCMT remains the production of high quality data in a safe and efficient manner. The on-sky productive time performance is seen to be slightly above what was recorded prior to Remote Observing and with no major faults, the observatory considers the switch over to Remote Observing a success. 

One unforeseen benefit from the switch to remote operations has been TSS in person interactions with science staff in the EAO Hilo building allowed for quicker feedback loops on project queries, data quality and other issues (at least prior to COVID-19, since which the bulk of JCMT staff have worked from home). 

For other facilities considering the switch to remote observing, reflecting on the process undertaken at the JCMT highlights some key components in place prior to the switch. In terms of facility and instrumentation safety and performance:

\begin{itemize}
    \item[-] The ability to ``soft test" remote systems  enables a smoother transition to full remote observing. In the case of JCMT this was possible utilizing EO and making use of other opportunities to test systems. 
    \item[-] Overhaul of mechanical, electrical, instrument and cryogenic systems to enable as much as possible to be remote powered, controlled and monitored. 
    \item[-] Enabling the instrument suite to be monitored remotely, reducing faults above and beyond existing systems in place for unexpected and unattended warm-ups.
\end{itemize}

The confidence observatory staff had in the continued scientific output after switching the remote operations can be attributed to:

\begin{itemize}
    \item[-] A Flexible Queue Observing Policy combined with comprehensive tools to communicate between PI and observatory.
    \item[-] TMU enabling optimal use of telescope time.
    \item[-] Ease of on-the-fly data quality assessment, by highly trained TSSs. with follow up by science support staff and users thanks to existing tools (ORAC-DR, OMP, JSA, JSA tracking pages).
\end{itemize}

\acknowledgments

The authors wish to acknowledge all current and past JCMT Extended Operators who have contributed to the success of the JCMT over the years including: Eldridge Shay, Cameron Wipper, E'lisa Lee, Kevin Silva, Alyssa Clark, Alexis Acohido, Callie Matulonis and Patrice Smith. In addition past JCMT staff and community members involved in the work and the vision of Remote Observing Ian Campbell, Tim Chuter, Iain Coulson, Henry Stilmack, Jim Hoge, Doug Johnstone, Gary Davis, Antonio Chrysostomou, Christine Wilson, and Walter Gear. 

The James Clerk Maxwell Telescope is operated by the East Asian Observatory on behalf of The National Astronomical Observatory of Japan; Academia Sinica Institute of Astronomy and Astrophysics; the Korea Astronomy and Space Science Institute; Center for Astronomical Mega-Science (as well as the National Key R$\&$D Program of China with No. 2017YFA0402700). Additional funding support is provided by the Science and Technology Facilities Council of the United Kingdom and participating universities in the United Kingdom and Canada. The James Clerk Maxwell Telescope has historically been operated by the Joint Astronomy Centre on behalf of the Science and Technology Facilities Council of the United Kingdom, the National Research Council of Canada and the Netherlands Organisation for Scientific Research. \Namakanui\ was constructed and funded by ASIAA in Taiwan, with funding for the mixers provided by ASIAA and at 230\,GHz by EAO. The \Namakanui\ instrument is a backup receiver for the GLT.

The authors wish to recognize and acknowledge the very significant cultural role and reverence that the summit of Maunakea has always had within the indigenous Hawaiian community.  We are most fortunate to have the opportunity to conduct astronomical observations from this mountain.

% References
\bibliography{main} % bibliography data in main.bib
\bibliographystyle{spiebib} % makes bibtex use spiebib.bst

\end{document}